\pdfoutput=1
\documentclass[runningheads,a4paper,orivec]{llncs}

\usepackage{vfpt}

\begin{document}

\title{A Higher-Order Abstract Syntax Approach to\\
  Verified$\;$Transformations$\;$on$\;$Functional$\;$Programs}
\titlerunning{Verified Transformations on Functional Programs}

\author{
  Yuting Wang
  \and
  Gopalan Nadathur
}
\authorrunning{Wang and Nadathur}
\tocauthor{
  Yuting Wang
  and Gopalan Nadathur (UMN)
}
\institute{
  University of Minnesota, USA\\
  \email{{\char"7B}yuting,gopalan{\char"7D}@cs.umn.edu}
}

\maketitle

\begin{abstract}

\noindent We describe an approach to the verified implementation
of transformations on functional programs that exploits the
higher-order representation of syntax. 
In this approach, transformations are specified using the logic of
hereditary Harrop formulas. 
On the one hand, these specifications serve directly as
implementations, being programs in the language \LProlog.
On the other hand, they can be used as input to the Abella system
which allows us to prove properties about them and thereby about the
implementations.  
We argue that this approach is especially effective in
realizing transformations that analyze binding structure. 
We do this by describing concise encodings in \LProlog for
transformations like typed closure conversion and code hoisting that
are sensitive to such structure and by showing how to prove their
correctness using Abella.
%
\end{abstract}

\section{Introduction}
\label{sec:intro}

This paper concerns the verification of compilers for functional
(programming) languages.
The interest in this topic is easily explained.
Functional languages support an abstract view of computation 
that makes it easier to construct programs and the resulting code also
has a flexible structure. 
Moreover, these languages have a strong mathematical basis that
simplifies the process of proving programs to be correct. 
However, there is a proviso to this observation: to derive the
mentioned benefit, the reasoning must be done relative to the abstract
model underlying the language, whereas programs are typically executed
only in their compiled form.
To close the gap, it is important also to ensure that the compiler
that carries out the translation preserves the meanings of programs.

The key role that compiler verification plays in overall program
correctness has been long recognized; e.g. see
\cite{mccarthy67ams,milner72mi} for early work on this topic.
With the availability of sophisticated systems such as
Coq~\cite{bertot04book}, Isabelle~\cite{nipkow02book} and
HOL~\cite{gordon91tphol} for mechanizing reasoning, impressive strides
have been taken in recent years towards actually verifying compilers for real
languages, as seen, for instance, in the CompCert project~\cite{leroy06popl}.
Much of this work has focused on compiling imperative
languages like C.
Features such as higher-order and nested functions that are present in
functional languages bring an additional complexity to their
implementation. 
A common approach to treating such features is to apply
transformations to programs that render them into a form to which
more traditional compilation methods can be applied.
These transformations must manipulate binding structure in
complex ways, an aspect that requires special
consideration at both the implementation and the verification
level~\cite{aydemir05tphols}.

Applications such as those above have motivated research 
towards developing good methods for representing and manipulating 
binding structure. 
Two particular approaches that have emerged from this work are those
that use the nameless representation of bound variables due to De
Bruijn~\cite{debruijn72} and the nominal logic framework of
Pitts~\cite{pitts03ic}.
These approaches provide an elegant treatment of aspects such as
$\alpha$-convertibility but do not directly support the analysis of 
binding structure or the realization of binding-sensitive operations
such as substitution.
A third approach, commonly known as the \emph{higher-order abstract
  syntax} or \HOAS approach, uses the abstraction operator in a typed
$\lambda$-calculus to represent binding structure in object-language
syntax. 
When such representations are embedded within a suitable logic, they
lead to a succinct and flexible treatment of many binding related
operations through $\beta$-conversion and unification.

The main thesis of this paper, shared with other work such as
\cite{hannan92lics} and \cite{belanger13cpp}, is that the \HOAS
approach is in fact well-adapted to the task of implementing and
verifying compiler transformations on functional languages.
Our specific objective is to demonstrate the usefulness of
a particular framework in this task.
This framework comprises two parts: the \LProlog
language~\cite{nadathur88iclp} that is implemented, for example, in
the Teyjus system~\cite{teyjus.website}, and the \Abella proof
assistant~\cite{baelde14jfr}.
The \LProlog language is a realization of the hereditary
Harrop formulas or \HOHH logic~\cite{miller12proghol}.
We show that this logic, which uses the simply typed
$\lambda$-calculus as a means for representing objects, is a suitable
vehicle for specifying transformations on functional programs.
Moreover, \HOHH specifications have a computational
interpretation that makes them \emph{implementations} of compiler
transformations.
The \Abella system is also based on a logic that supports the \HOAS
approach.
This logic, which is called \Gee, incorporates a treatment of
fixed-point definitions that can also be interpreted inductively or
co-inductively. 
The \Abella system uses these definitions to embed \HOHH within \Gee
and thereby to reason directly about the specifications written in
\HOHH.
As we show in this paper, this yields a convenient means for verifying 
implementations of compiler transformations.

An important property of the framework that we consider, as also of
systems like LF~\cite{harper93jacm} and Beluga~\cite{pientka10ijcar},
is that it uses a weak $\lambda$-calculus for representing objects.
There have been attempts to derive similar benefits from using 
functional languages or the language underlying systems such as Coq. 
Some benefits, such as the correct implementation of substitution, can
be obtained even in these contexts.
However, the equality relation embodied in these systems is very
strong and the analysis of $\lambda$-terms in them is therefore not
limited to examining just their syntactic structure.
This is a significant drawback, given that such examination plays a
key role in the benefits we describe in this paper. 
In light of this distinction, we shall use the term
\emph{$\lambda$-tree   syntax} \cite{miller00cl} for the more
restricted  version of \HOAS whose use is the focus of our
discussions. 

The rest of this paper is organized as follows. In
Section~\ref{sec:framework} we introduce the reader to the framework
mentioned above. We then show in succeeding sections how this
framework can be used to implement and to verify transformations 
on functional programs. We conclude the paper by discussing 
the relationship of the ideas we describe here to other
existing work.\footnote{The actual development of several of
  the proofs discussed in this paper can be found at the URL
  \url{http://www-users.cs.umn.edu/~gopalan/papers/compilation/}.}

\section{The Framework}
\label{sec:framework}

\vspace{-0.1cm}
We describe, in turn, the specification logic and \LProlog, the
reasoning logic, and the manner in which the Abella system embeds the
specification logic.

\vspace{-0.3cm}
\subsection{The specification logic and \LProlog}

The \HOHH logic is an intuitionistic and predicative fragment of
Church's Simple Theory of Types~\cite{church40}.
Its types are formed using the function type
constructor \lsti|->| over user defined primitive types and the
distinguished type \lsti|o| for formulas. Expressions are formed from
a user-defined \emph{signature} of typed constants
whose argument types do not contain \lsti|o|
and the \emph{logical constants} $\simply$ and $\sconj$ of type
\lsti|o -> o -> o| and $\Pi_\tau$ of type \lsti|($\tau$ -> o) -> o|
for each type $\tau$ not containing \lsti|o|. We write $\simply$ and
$\sconj$, which denote implication and conjunction respectively, in
infix form. Further, we write $\Pi_\tau\app \lambda (x:\tau) M$, which
represents the universal quantification of $x$ over $M$, as
$\typedforall{\tau}{x}{M}$.

The logic is oriented around two sets of formulas called
\emph{goal formulas} and \emph{program clauses} that are given by the
following syntax rules:
\vspace{-0.15cm}
\begin{tabbing}
\qquad\=$G$\qquad\=::=\qquad\=\kill
\>$G$\>::=\>$A\sep G \;\sconj\; G\sep D\simply G \sep \typedforall{\tau}{x}{G}$\\
\>$D$\>::=\>$A\sep G\simply A\sep \typedforall{\tau}{x}{D}$
\end{tabbing}
\vspace{-0.15cm}
Here, $A$ represents \emph{atomic formulas} that have
the form $(p\app t_1\app \ldots\app t_n)$ where $p$ is a (user
defined) \emph{predicate constant}, \ie a constant with target type
\lsti|o|.
Goal formulas of the last two kinds are referred to as hypothetical
and universal goals.
Using the notation $\Pi_{\bar{\tau}}\bar{x}$ to denote a
sequence of quantifications, we see that a program clause has the form
$\typedforall{\bar{\tau}}{\bar{x}}{A}$ or
$\typedforall{\bar{\tau}}{\bar{x}}{G \simply A}$. We refer to $A$ as
the head of such a clause and $G$ as the body; in the first case the
body is empty.

A collection of program clauses constitutes a \emph{program}.
A program and a signature represent a specification of all
the goal formulas that can be derived from them.
The derivability of a goal formula $G$ is expressed formally by the
judgment $\sequent{\Sigma}{\Theta}{\Gamma}{G}$ in which $\Sigma$ is a
signature, $\Theta$ is a collection of program clauses defined by the
user and $\Gamma$ is a collection of dynamically added program
clauses.
The validity of such a judgment---also called a sequent---is
determined by provability in intuitionistic logic but can equivalently
be characterized in a goal-directed fashion as follows.
If $G$ is conjunctive, it yields sequents for ``solving''
each of its conjuncts in the obvious way. If it is a hypothetical or a
universal goal, then one of the following rules is used:

\vspace{-0.1cm}
\begin{smallgather}
  \infer[\impR]
    {\sequent{\Sigma}{\Theta}{\Gamma}{D \simply G}}
    {\sequent{\Sigma}{\Theta}{\Gamma, D}{G}}
  \qquad
  \infer[\forallR]
     {\sequent{\Sigma}{\Theta}{\Gamma}{\typedforall{\tau}{x}{G}}}
     {(c \notin \Si) & \sequent{\Sigma,c:\tau}{\Theta}{\Gamma}{G[c/x]}}
\end{smallgather}

\vspace{-0.1cm}
\noindent In the \forallR\ rule, $c$ must be a constant not already
in $\Sigma$; thus, these rules respectively cause the program
and the signature to grow while searching for a derivation.
Once $G$ has been simplified to an atomic formula, the
sequent is derived by generating an instance of a clause from $\Theta$
or $\Gamma$ whose head is identical to $G$ and by constructing a
derivation of the corresponding body of the clause if it is
non-empty. This operation is referred to as backchaining on a
clause.

In presenting \HOHH specifications in this paper
we will show programs as a sequence of clauses each terminated by a
period.
We will leave the outermost universal quantification in these clauses
implicit, indicating the variables they bind by using tokens that
begin with uppercase letters.
We will write program clauses of the form  $G \simply A$ as
$A$~\lsti+:-+$G$. We will show goals of the form $G_1 \land G_2$ and
$\typedforall{\tau}{y}{G}$ as $G_1$\lsti+,+$G_2$ and
\lsti+pi+~$y:\tau$\lsti+\+~$G$, respectively, dropping the type
annotation in the latter if it can be filled in uniquely based on the
context.  Finally, we will write abstractions as $y$\lsti+\+$M$
instead of $\lambdax{y}{M}$.

Program clauses provide a natural means for encoding rule based specifications.
Each rule translates into a clause whose head corresponds to the
conclusion and whose body represents the premises of the rule.
These clauses embody additional mechanisms that simplify the treatment
of binding structure in object languages.
They provide $\lambda$-terms as a means for representing objects,
thereby allowing binding to be reflected into an explicit
meta-language abstraction.
Moreover, recursion over such structure, that is typically treated
via side conditions on rules
expressing requirements such as freshness for variables,
can be captured precisely through universal and hypothetical goals.
This kind of encoding is concise and has logical properties that we
can use in reasoning.

We illustrate the above ideas by considering the specification of the
typing relation for the simply typed $\lambda$-calculus (STLC).
Let $N$ be the only atomic type.
We use the \HOHH type \lsti+ty+ for representations of object language
types that we build using the constants \lsti+n : ty+ and
\lsti+arr : ty -> ty -> ty+.
Similarly, we use the \HOHH type \lsti+tm+ for encodings of object language
terms  that we build using the
constants  \lsti+app : tm -> tm -> tm+ and
\lsti+abs : ty -> (tm -> tm) -> tm+.
The type of the latter constructor follows our chosen
approach to encoding binding: for example, we represent the \STLC
expression $(\typedlambda{y}{N \to N}{\typedlambda{x}{N}{(y \app
x)}})$ by the \HOHH term
\lsti+(abs (arr n n) (y\ (abs n (x\ (app y x)))))+.
Typing for the STLC is a judgment written as $\Gamma \tseq T :
\mbox{\it Ty}$ that expresses a relationship between a context
$\Gamma$ that assigns types to variables, a term $T$ and a type
$\mbox{\it Ty}$. Such judgments are derived using the following rules:
\vspace{-0.1cm}
\begin{smallgather}
  \infer[]{
    \Gamma \tseq T_1 \app T_2 : \mbox{\it Ty}_2
  }{
    \Gamma \tseq T_1 : \mbox{\it Ty}_1 \to \mbox{\it Ty}_2
    &
    \Gamma \tseq T_2 : \mbox{\it Ty}_1
  }
  \quad
  \infer[]{
    \Gamma \tseq \typedlambda{y}{\mbox{\it Ty}_1}T : (\mbox{\it Ty}_1 \to \mbox{\it Ty}_2)
  }{
    \Gamma, y:\mbox{\it Ty}_1 \tseq T : \mbox{\it Ty}_2
  }
\end{smallgather}
\vspace{-0.5cm}

The second rule has a proviso: $y$ must be fresh to $\Gamma$.
In the $\lambda$-tree syntax approach, we encode typing as a
binary relation between a term and a type, treating the typing context
implicitly via dynamically added clauses.
Using the predicate \lsti|of| to represent this relation, we define it
through the following clauses:

\vspace{-0.10cm}
\begin{lstlisting}
of (app T1 T2) Ty2 :- of T1 (arr Ty1 Ty2), of T2 Ty1.
of (abs Ty1 T) (arr Ty1 Ty2) :-  pi y\ (of y Ty1 => of (T y) Ty2).
\end{lstlisting}
\vspace{-0.10cm}

\noindent The second clause effectively says that
\lsti|(abs Ty1 T)| has the type \lsti|(arr Ty1 Ty2)| if \lsti|(T y)|
has type \lsti|Ty2| in an extended context that assigns \lsti|y| the
type \lsti|Ty1|. Note that the universal goal ensures that \lsti|y| is
new and, given our encoding of terms, \lsti|(T y)| represents the body
of the object language abstraction in which the bound variable has
been replaced by this new name.

The rules for deriving goal formulas give \HOHH specifications a
computational interpretation.
We may also leave particular parts of a goal unspecified, representing
them by ``meta-variables,'' with the intention that values be found for
them that make the overall goal derivable.
This idea underlies the language $\lambda$Prolog that is implemented,
for example, in the Teyjus system~\cite{teyjus.website}.

\vspace{-0.3cm}
\subsection{The reasoning logic and Abella}\label{sec:reasoning}

The inference rules that describe a relation are usually meant to be
understood in an ``if and only if'' manner.
Only the ``if'' interpretation is relevant to using rules to effect
computations and their encoding in the \HOHH logic captures this part
adequately.
To reason about the \emph{properties} of the resulting computations,
however, we must formalize the ``only if'' interpretation as well.
This functionality is realized by the logic \Gee that is implemented in
the Abella system.

The logic \Gee is also based on an intuitionistic and predicative
version of Church's Simple Theory of Types.
Its types are like those in \HOHH except that the type \lsti|prop|
replaces \lsti|o|.
Terms are formed from user-defined constants whose argument types do
not include \lsti|prop| and the following logical constants:
\lsti|true| and \lsti|false| of type \lsti|prop|; $\conj$, ${\lor}$
and \lsti|->| of type \lsti|prop -> prop -> prop| for conjunction,
disjunction and implication; and, for every type $\tau$ not containing
\lsti|prop|, the quantifiers $\forall_\tau$ and $\exists_\tau$ of type
\lsti|($\tau$ -> prop) -> prop| and the equality symbol
\lsti|=$_\tau$| of type \lsti|$\tau$ -> $\tau$ -> prop|.
The formula $B =_\tau B'$ holds if and only if $B$ and $B'$ are of
type $\tau$ and equal under $\alpha\beta\eta$ conversion. We will omit
the type $\tau$ in logical constants when its identity is clear from
the context.

A novelty of \Gee is that it is parameterized by \emph{fixed-point
  definitions}.
Such definitions consist of a collection of \emph{definitional clauses}
each of which has the form
$\forall {\bar{x}}, A \defeq B$ where $A$ is an atomic formula all of
whose free variables are bound by $\bar{x}$ and $B$ is a formula whose
free variables must occur in $A$; $A$ is called the head of such
a clause and $B$ is called its body.\footnote{To be acceptable,
  definitions must cumulatively satisfy certain stratification
  conditions~\cite{mcdowell00tcs} that we adhere to in the paper but
do not explicitly discuss.}
To illustrate definitions, let \lsti|olist| represent the type of
lists of \HOHH formulas and let \lsti|nil| and \lsti|::|, written in
infix form, be constants for building such lists. Then the append
relation at the \lsti|olist| type is defined in \Gee by the following clauses:
\vspace{-0.1cm}
\begin{lstlisting}
append nil L L;
append (X :: L1) L2 (X :: L3) := append L1 L2 L3.
\end{lstlisting}
\vspace{-0.1cm}
This presentation also illustrates several conventions used in writing
definitions: clauses
of the form $\forall {\bar{x}}, A \defeq\,$\lsti|true| are abbreviated to
$\forall {\bar{x}}, A$, the outermost universal quantifiers in a clause
are made implicit by representing the variables they bind by tokens
that start with an uppercase letter, and a sequence of clauses is
written using semicolon as a separator and period as a terminator.

The proof system underlying \Gee interprets atomic
formulas via the fixed-point definitions.
Concretely, this means that definitional clauses can be used in two
ways.
First, they may be used in a backchaining mode to
derive atomic formulas: the formula is matched
with the head of a clause and the task is reduced to deriving the
corresponding body.
Second, they can also be used to do case analysis on an assumption. Here
the reasoning structure is that if an atomic formula holds, then it
must be because the body of one of the clauses
defining it holds. It therefore suffices to show that the
conclusion follows from each such possibility.

The clauses defining a particular predicate can further be interpreted
inductively or coinductively, leading to corresponding reasoning
principles relative to that predicate. As an
example of how this works, consider proving
\vspace{-0.1cm}
\begin{lstlisting}
forall L1 L2 L3, append L1 L2 L3 -> append L1 L2 L3' -> L3 = L3'
\end{lstlisting}
\vspace{-0.1cm}
assuming that we have designated \lsti|append| as an inductive
predicate. An induction on the first occurrence of
\lsti|append| then allows us to assume that the entire formula holds
any time the leftmost atomic formula is replaced by a
formula that is obtained by unfolding its definition and that has
\lsti|append| as its predicate head.

Many arguments concerning binding require the capability
of reasoning over structures with free variables where each such
variable is treated as being distinct and not further analyzable.
To provide this capability, \Gee includes the special \emph{generic}
quantifier $\nabla_\tau$, pronounced as ``nabla'', for each type
$\tau$ not containing \lsti|prop|~\cite{miller05tocl}.
In writing this quantifier, we, once again, elide the type
$\tau$.
The rules for treating $\nabla$ in an assumed formula and a
formula in the conclusion are similar: a ``goal'' with
\lsti|(|$\nabla$\lsti|x M)| in it reduces to one in which this formula
has been replaced by \lsti|M[c/x]| where \lsti|c| is a fresh,
unanalyzable constant called a \emph{nominal constant}.
Note that \lsti|nabla| has a meaning that is different from that of
\lsti|forall|: for example, \lsti|(nabla x y, x = y -> false)| is
provable but \lsti|(forall x y, x = y -> false)| is not.

\Gee allows the \lsti|nabla| quantifier to be used also in the heads
of definitions. The full form for a
definitional clause is in fact $\forall {\bar{x}} \nabla {\bar{z}}, A
\defeq B$, where the \lsti|nabla| quantifiers scope only over $A$. In
generating an instance of such a clause, the variables in $\bar{z}$
must be replaced with nominal constants. The quantification order
then means that the instantiations of the variables in $\bar{x}$
cannot contain the constants used for $\bar{z}$. This extension makes
it possible to encode structural properties of terms in
definitions. For example, the clause
\lsti|(nabla x, name x)| defines \lsti|name| to be a
recognizer of nominal constants. Similarly, the
clause \lsti|(nabla x, fresh x B)| defines \lsti|fresh| such
that  \lsti|(fresh X B)| holds just in the case that
\lsti|X| is a nominal constant and \lsti|B| is a term that does not
contain \lsti|X|.
As a final example, consider the following clauses in which
\lsti|of| is the typing predicate from the previous subsection.
\vspace{-0.1cm}
\begin{lstlisting}
ctx nil;
nabla x, ctx (of x T :: L) :=  ctx L.
\end{lstlisting}
\vspace{-0.1cm}
These clauses define \lsti|ctx| such that \lsti|(ctx L)| holds exactly
when \lsti|L| is a list of type assignments to distinct variables.

\vspace{-0.3cm}
\subsection{The two-level logic approach}\label{sec:twolevel}

Our framework allows us to write specifications in \HOHH and reason
about them using \Gee.
Abella supports this \emph{two-level logic approach} by encoding \HOHH
derivability in a definition and providing a convenient interface to
it.
The user program and signature for these derivations are obtained
from a \LProlog program file.
The state in a derivation is represented by a judgment of the
form \lsti+{$\Gamma$ |- $G$}+ where $\Gamma$ is the list of
dynamically added clauses; additions to the signature are treated
implicitly via nominal constants.
If $\Gamma$ is empty, the judgment is abbreviated to
\lsti|{G}|.
The theorems that are to be proved mix such judgments with other ones
defined directly in Abella.
For example, the uniqueness of typing for the
STLC based on its encoding in \HOHH can be stated as follows:
\vspace{-0.1cm}
\begin{lstlisting}
forall L M T T', ctx L -> {L |- of M T} -> {L |- of M T'} -> T = T'.
\end{lstlisting}
\vspace{-0.1cm}
This formula talks about the typing of \emph{open} terms
relative to a dynamic collection of clauses that assign unique types
to (potentially) free variables.

The ability to mix specifications in \HOHH and definitions in Abella
provides considerable expressivity to the reasoning process.
This expressivity is further enhanced by the fact that both \HOHH and
\Gee support the $\lambda$-tree syntax approach.
We illustrate these observations by considering the explicit treatment
of substitutions.
We use the type \lsti|map| and the constant \lsti|map: tm -> tm -> map|
to represent mappings for individual variables (encoded as nominal
constants) and a list of such
mappings to represent a substitution; for simplicity, we overload
the constructors \lsti|nil| and \lsti|::| at this type.
Then the predicate \lsti|subst| such that \lsti|subst ML M M'| holds
exactly when \lsti|M'| is the result of applying the substitution
\lsti|ML| to \lsti|M| can be defined by the following clauses:
\vspace{-0.1cm}
\begin{lstlisting}
subst nil M M;
nabla x, subst ((map x V) :: ML) (R x) M := subst ML (R V) M.
\end{lstlisting}
\vspace{-0.1cm}
Observe how quantifier ordering is used in this definition to
create a ``hole'' where a free variable appears in a term and
application is then used to plug the hole with the substitution.
This definition makes it extremely easy to prove structural properties
of substitutions.
For example, the fact that substitution distributes over applications
and abstractions can be stated as follows:
\vspace{-0.1cm}
\begin{lstlisting}
forall ML M1 M2 M', subst ML (app M1 M2) M' ->
   exists M1' M2', M' = app M1' M2' /\  subst ML M1 M1' /\ subst ML M2 M2'.
forall ML R T M', subst ML (abs T R) M' ->
   exists R', M' = abs T R' /\ nabla x, subst ML (R x) (R' x).
\end{lstlisting}
\vspace{-0.1cm}
An easy induction over the definition of substitution proves these properties.

As another example, we may want to characterize relationships between
closed terms and substitutions.
For this, we can first define well-formed terms through the following
\HOHH clauses:
\vspace{-0.1cm}
\begin{lstlisting}
tm (app M N) :- tm M, tm N.
tm (abs T R) :- pi x\ tm x => tm (R x).
\end{lstlisting}
\vspace{-0.1cm}
Then we characterize the context used in \lsti|tm| derivations in Abella
as follows:
\vspace{-0.1cm}
\begin{lstlisting}
tm_ctx nil;
nabla x, tm_ctx (tm x :: L) := tm_ctx L.
\end{lstlisting}
\vspace{-0.1cm}
Intuitively, if \lsti|tm_ctx L| and \lsti+{L |- tm M}+ hold, then \lsti|M| is a
well-formed term whose free variables are given by \lsti|L|.
Clearly, if \lsti+{tm M}+ holds, then \lsti|M| is closed.
Now we can state the fact that  a closed term is unaffected by a
substitution:
\vspace{-0.1cm}
\begin{lstlisting}
forall ML M M', {tm M} -> subst ML M M' -> M = M'.
\end{lstlisting}
\vspace{-0.1cm}
Again, an easy induction on the definition of substitutions proves
this property.

\section{Implementing Transformations on Functional Programs}
\label{sec:implfpt}

We now turn to the main theme of the paper, that of showing the
benefits of our framework in the verified implementation of
compilation-oriented program transformations for functional
languages.
The case we make has the following broad structure.
Program transformations are often conveniently described in a
syntax-directed and rule-based fashion.
Such descriptions can be encoded naturally using the program clauses
of the \HOHH logic.
In transforming functional programs, special attention must be paid to
binding structure.
The $\lambda$-tree syntax approach, which is supported by the
\HOHH logic, provides a succinct and logically precise means for
treating this aspect.
The executability of \HOHH specifications renders them immediately
into implementations.
Moreover, the logical character of the specifications is
useful in the process of reasoning about their correctness.

This section is devoted to substantiating our claim concerning
implementation.
We do this by showing how to specify transformations that are
used in the compilation of functional languages.
An example we consider in detail is that of closure conversion.
Our interest in this transformation is twofold.
First, it is an important step in the compilation of functional
programs: it is, in fact, an enabler for other transformations such as
code hoisting.
Second, it is a transformation that involves a complex manipulation of
binding structure.
Thus, the consideration of this transformation helps shine a light on
the special features of our framework.
The observations we make in the context of closure conversion are
actually applicable quite generally to the compilation process.
We close the section by highlighting this fact relative to other
transformations that are of interest.

\vspace{-0.3cm}
\subsection{The closure conversion transformation}
\label{ssec:cc}

The closure conversion transformation is designed to replace (possibly
nested) functions in a program by \emph{closures} that each consist of
a function and an environment.
The function part is obtained from the original function by replacing
its free variables by projections from a new environment parameter.
Complementing this, the environment component encodes the construction 
of a value for the new parameter in the enclosing context.
For example, when this transformation is applied to the
following pseudo OCaml code segment

\vspace{-0.1cm}
\begin{lstlisting}[language=Caml]
let x = 2 in let y = 3 in (fun z. z + x + y)
\end{lstlisting}
\vspace{-0.1cm}
it will yield
\vspace{-0.1cm}
\begin{lstlisting}[language=Caml]
let x = 2 in let y = 3 in <fun z e. z$\;$+$\;$e.1$\;$+$\;$e.2, (x,y)>
\end{lstlisting}
\vspace{-0.1cm}
We write \lsti|<F,E>| here to represent a closure whose
function part is \lsti|F| and environment part is \lsti|E|, and
\lsti|e.i| to represent the $i$-th projection applied to an
``environment parameter'' \lsti|e|. This transformation makes the
function part independent of the context in which it appears, thereby
allowing it to be extracted out to the top-level of the program.

\vspace{-0.3cm}
\subsubsection{The source and target languages.}
\label{sssec:cclangs}

\begin{figure}[t]
  \centering
  \parbox{0.4\linewidth}{
    \begin{smallalign}
      & T ::= \tnat \sep T_1 \to T_2 \sep \tunit \sep  {T_1} \tprod {T_2}\\
      & M ::= n \sep x \sep \pred M \sep M_1 + M_2\\
      & \qquad \sep \ifz {M_1} {M_2} {M_3} \\
      & \qquad \sep \unit \sep \pair {M_1} {M_2} \sep \fst M \sep \snd M\\
      & \qquad \sep \letexp x {M_1} {M_2} \\
      & \qquad \sep \fix f x M \sep (M_1 \app M_2)\\
      & V ::= n \sep \fix f x M \sep () \sep \pair {V_1} {V_2}
    \end{smallalign}
\vspace{-0.6cm}
    \caption{Source language syntax}
    \label{fig:sourcelang}
  }
  \qquad
  \parbox{0.4\linewidth}{
    \begin{smallalign}
      & T ::= \tnat \sep {T_1} \to {T_2} \sep {T_1} \carr {T_2} \sep \tunit \sep  {T_1} \tprod {T_2}\\
      & M ::= n \sep x \sep \pred M \sep M_1 + M_2 \\
      & \qquad \sep \ifz {M_1} {M_2} {M_3} \\
      & \qquad \sep \unit \sep \pair {M_1} {M_2} \sep \fst M \sep \snd M\\
      & \qquad \sep \letexp x {M_1} {M_2} \sep \ \abs x M \sep (M_1 \app M_2)\\
      & \qquad \sep \clos{M_1}{M_2} \sep \open {x_f} {x_e} {M_1} {M_2}\\
      & V ::= n \sep \abs x M \sep () \sep \pair {V_1} {V_2} \sep \clos {V_1} {V_2}
    \end{smallalign}
\vspace{-0.6cm}
    \caption{Target language syntax}
    \label{fig:targlang}
  }
\vspace{-0.6cm}
\end{figure}

Figures~\ref{fig:sourcelang} and \ref{fig:targlang} present the
syntax of the source and target languages that we shall use in this
illustration.
In these figures, $T$, $M$ and $V$ stand respectively for the
categories of types, terms and the terms recognized as values. $\tnat$
is the type for natural numbers and $n$ corresponds to constants of
this type.
Our languages include some arithmetic operators, the conditional and
the tuple constructor and destructors; note that $\predsans$ represents the
predecessor function on numbers, the behavior of the conditional is
based on whether or not the ``condition'' is zero and $\fstsans$ and
$\sndsans$ are the projection operators on pairs.
The source language includes the recursion operator $\fixsans$ which
abstracts simultaneously over the function and the parameter; the
usual abstraction is a degenerate case in which the function parameter does
not appear in the body.
The target language includes the expressions $\clos {M_1} {M_2}$ and
$(\open {x_f} {x_e} {M_1} {M_2})$ representing the formation and
application of closures. 
The target language does not have an explicit fixed point constructor.
Instead, recursion is realized by parameterizing the function part of
a closure with a function component; this treatment should become
clear from the rules for typing closures and for evaluating the
application of closures that we present below.
The usual forms of abstraction and application are included in the
target language to simplify the presentation of the transformation.
The usual function type is reserved for closures; abstractions are
given the type ${T_1} \carr {T_2}$ in the target language.
We abbreviate $\pair {M_1} {\ldots, \pair {M_n} \unit}$ by $(M_1,\ldots,M_n)$
and $\fst {(\snd {(\ldots(\snd M))})}$ where $\mathbf{snd}$ is applied
$i-1$ times for $i \geq 1$ by $\pi_i(M)$.

Typing judgments for both the source and target languages are written
as $\Gamma \tseq M : T$, where $\Gamma$ is a list of type assignments
for variables.
The rules for deriving typing judgments are routine, with the
exception of those for introducing and eliminating closures in the
target language that are shown below:
\begin{smallgather}
  \infer[\cofclos]{
    \Gamma \tseq {\clos {M_1} {M_2}} : {T_1 \to T_2}
  }{
    \tseq {M_1} : {((T_1 \to T_2) \tprod T_1 \tprod T_e) \carr T_2}
    &
    \Gamma \tseq {M_2} : {T_e}
  }
  \\[5pt]
  \infer[\cofopen]{
    \Gamma \tseq  {\open {x_f} {x_e} {M_1} {M_2}} : T
  }{
    \Gamma \tseq {M_1} : {T_1 \to T_2}
    &
    {\Gamma, x_f:((T_1 \to T_2) \tprod T_1 \tprod l) \carr T_2, x_e:l} \tseq {M_2} : T
  }
\end{smallgather}
In $\cofclos$, the function part of a closure must be typable in an
empty context.
In $\cofopen$, $x_f$, $x_e$ must be names that are new to $\Gamma$.
This rule also uses a ``type'' $l$ whose meaning must be explained.
This symbol represents a new type constant, different from $\tnat$ and
$\unit$ and any other type constant used in the typing derivation.
This constraint in effect captures the requirement that the
environment of a closure should be opaque to its user.

The operational semantics for both the source and the target language
is based on a left to right, call-by-value evaluation strategy.
We assume that this is given in small-step form and, overloading
notation again, we write $M \step{1} M'$ to denote that $M$ evaluates to
$M'$ in one step in whichever language is under consideration.
The only evaluation rules that may be non-obvious are the ones for
applications.
For the source language, they are the following:
\begin{smallgather}
  \infer[]{
    M_1 \app M_2 \step{1} M_1' \app M_2
  }{
    M_1 \step{1} M_1'
  }
  \quad\
  \infer[]{
    V_1 \app M_2 \step{1} V_1 \app M_2'
  }{
    M_2 \step{1} M_2'
  }
\quad\
  \infer[]{
    (\fix f x M) \app V \step{1} M[\fix f x M/f, V/x]
  }{}
\end{smallgather}
For the target language, they are the following:
\begin{smallgather}
  \infer[]{
    {\open {x_f} {x_e} {M_1} {M_2}} \step{1} {\open {x_f} {x_e} {M_1'} {M_2}}
  }{
    {M_1} \step{1} {M_1'}
  }
  \\[7pt]
  \infer[]{
    {\open {x_f} {x_e} {\clos {V_f} {V_e}} {M_2}} \step{1}
    {M_2[V_f/x_f, V_e/x_e]}
  }{}
\end{smallgather}
One-step evaluation generalizes in the obvious way to $n$-step
evaluation that we denote by $M \step{n} M'$.
Finally, we write $M \eval V$ to denote the evaluation of $M$ to the
value $V$ through $0$ or more steps.

\vspace{-0.4cm}
\subsubsection{The transformation.}
\label{sssec:cctrans}

\begin{figure*}[!t]
\begin{smallgather}
  \infer[\ccnat]{
    \cc {\rho} n n
  }{}
  \qquad
  \infer[\ccvar]{
    \cc {\rho} x M
  }{
    (x \mapsto M) \in \rho
  }
  \qquad
  \infer[\ccfvs]{
    \ccenv {\rho} {(x_1,\ldots,x_n)} {(M_1,\ldots,M_n)}
  }{
    \cc {\rho} {x_1} {M_1}
    &
    \ldots
    &
    \cc {\rho} {x_n} {M_n}
  }
  \\
  \infer[\ccpred]{
    \cc {\rho} {\pred M} {\pred M'}
  }{
    \cc{\rho} M {M'}
  }
  \qquad
  \infer[\ccplus]{
    \cc {\rho} {M_1 + M_2} {M_1' + M_2'}
  }{
    \cc \rho {M_1} {M_1'}
    &
    \cc \rho {M_2} {M_2'}
  }
  \\
  \infer[\ccifz]{
    \cc {\rho} {\ifz {M} {M_1} {M_2}} {\ifz {M'} {M_1'} {M_2'}}
  }{
    \cc \rho {M} {M'}
    &
    \cc \rho {M_1} {M_1'}
    &
    \cc \rho {M_2} {M_2'}
  }
  \qquad
  \infer[\ccunit]{
    \cc \rho \unit \unit
  }{}
  \\
  \infer[\ccpair]{
    \cc {\rho} {\pair {M_1} {M_2}} {\pair {M_1'} {M_2'}}
  }{
    \cc \rho {M_1} {M_1'}
    &
    \cc \rho {M_2} {M_2'}
  }
  \quad
  \infer[\ccfst]{
    \cc {\rho} {\fst M} {\fst M'}
  }{
    \cc{\rho} M {M'}
  }
  \quad
  \infer[\ccsnd]{
    \cc {\rho} {\snd M} {\snd M'}
  }{
    \cc{\rho} M {M'}
  }
  \\
  \infer[\cclet\quad y\ \mbox{\rm must be fresh} ]{
    \cc \rho {\letexp x {M_1} {M_2}} {\letexp y {M_1'} {M_2'}}
  }{
    \cc \rho {M_1} {M_1'}
    &
    \cc {\rho, x \mapsto y} {M_2} {M_2'}
  }
  \\
  \infer[\ccapp\quad g\ \mbox{\rm must be fresh}]{
    \cc {\rho} {M_1 \app M_2}
        {\letexp g {M_1'}
          {\open {x_f} {x_e} {g} {x_f \app (g,M_2',x_e)}}}
  }{
    \cc {\rho} {M_1} {M_1'}
    &
    \cc {\rho} {M_2} {M_2'}
  }
  \\
  \infer[\ccfix]{
    \cc {\rho} {\fix f x M}
        {\clos
          {\abs p {\letexp g {\pi_1(p)}
                   {\letexp y {\pi_2(p)}
                    {\letexp {x_e} {\pi_3(p)} {M'}}}}}
          {M_e}}
  }{
    (x_1,\ldots,x_n) \supseteq \fvars {\fix f x M}
    &
    \ccenv {\rho} {(x_1,\ldots,x_n)} {M_e}
    &
    \cc {\rho'} M {M'}
  }
  \\
  \mbox{where $\rho' = (x \mapsto y, f \mapsto g, x_1 \mapsto
    \pi_1(x_e), \ldots, x_n \mapsto \pi_n(x_e))$ and $p, g, y,$ and $x_e$ are
    fresh variables}
\end{smallgather}
\vspace{-0.5cm}
\caption{Closure Conversion Rules}
\label{fig:cc}
\vspace{-0.5cm}
\end{figure*}
In the general case, we must transform terms under
mappings for their free variables:
for a function term, this mapping represents the replacement of
the free variables by projections from the environment variable for
which a new abstraction will be introduced into the term.
Accordingly, we specify the transformation as a 3-place relation
written as $\cc \rho M {M'}$, where $M$ and $M'$ are source and target
language terms and $\rho$ is a mapping from (distinct) source language
variables to target language terms.
We write $(\rho, x \mapsto M)$ to denote the extension of $\rho$ with a
mapping for $x$ and $(x \mapsto M) \in \rho$ to mean that $\rho$
contains a mapping of $x$ to $M$.
Figure~\ref{fig:cc} defines the $\cc \rho M {M'}$ relation in a
rule-based fashion; these rules use the auxiliary relation $\ccenv
\rho {(x_1,\ldots,x_n)} {M_e}$ that determines an environment
corresponding to a tuple of variables.
The $\cclet$ and $\ccfix$ rules have a proviso: the bound variables,
$x$ and $f, x$ respectively, should have been renamed to avoid clashes
with the domain of $\rho$.
Most of the rules have an obvious structure. We comment only on the
ones for transforming fixed point expressions and applications.
The former translates into a closure.
The function part of the closure is obtained by transforming the body
of the abstraction, but under a new mapping for its free variables;
the expression $(x_1,\ldots,x_n) \supseteq \fvars{\fix f x M}$
means that all the free variables of $(\fix f x M)$ appear in the tuple.
The environment part of the closure correspondingly contains
mappings for the variables in the tuple that are determined by the
enclosing context.
Note also that the parameter for the function part of the closure is
expected to be a triple, the first item of which corresponds to the
function being defined recursively in the source language expression.
The transformation of a source language application makes clear how
this structure is used to realize recursion: the
constructed closure application has the effect of feeding the closure
to its function part as the first component of its argument.

\vspace{-0.3cm}
\subsection{A \LProlog rendition of closure conversion}
\label{ssec:implcc}

Our presentation of the implementation of closure conversion has two
parts: we first show how to encode the source and target languages and
we then present a \LProlog specification of the transformation. In the
first part, we discuss also the formalization of the evaluation and
typing relations; these will be used in the correctness proofs that we
develop later.

\vspace{-0.4cm}
\subsubsection{Encoding the languages.}
\label{sssec:encode_cclang}

We first consider the encoding of types. We will use \lsti|ty| as
the \LProlog type for this encoding for both languages. The constructors
\lsti|tnat|, \lsti|tunit| and \lsti|prod| will encode, respectively,
the natural number, unit and pair types. There are two arrow types to
be treated. We will represent $\to$ by \lsti|arr| and $\carr$ by
\lsti|arr'|. The following signature summarizes these decisions.

\vspace{-0.1cm}
{\footnotesize
\begin{tabbing}
\quad\=\lsti|tnat,tunit|\ \=\ \,\=\lsti|ty|\qquad\qquad\qquad\=\lsti|arr,prod,arr'|\ \=\ \,\=\kill
\>\lsti|tnat,tunit|    \>:\>  \lsti|ty| \>\lsti|arr,prod,arr'| \>:\>  \lsti|ty -> ty -> ty|
\end{tabbing}
}
\vspace{-0.1cm}

We will use the \LProlog type \lsti|tm| for encodings of source
language terms. The particular constructors that we will use for
representing the terms themselves are the following, assuming that
\lsti|nat| is a type for representations of natural numbers:

\vspace{-0.1cm}
{\footnotesize
\begin{tabbing}
\quad\=\lsti|plus,pair,app|\ :\ \lsti|tm -> tm -> tm|\quad\=\lsti|fix|\ :\ \lsti|(tm -> tm -> tm) -> tm|\quad\=\kill
\>\lsti|nat|\ :\ \lsti|nat -> tm|\>
\lsti|pred,fst,snd|\ :\ \lsti|tm -> tm|\>
\lsti|unit|\ :\ \lsti|tm|\\
\>\lsti|plus,pair,app|\ :\ \lsti|tm -> tm -> tm|\>
\lsti|ifz|\ :\ \lsti|tm -> tm -> tm -> tm|\\
\>\lsti|let|\ :\ \lsti|tm -> (tm -> tm) -> tm|\>
\lsti|fix|\ :\ \lsti|(tm -> tm -> tm) -> tm|
\end{tabbing}
}
\vspace{-0.1cm}
\noindent The only constructors that need further
explanation here are \lsti|let| and \lsti|fix|.
These encode binding constructs in the source language and, as
expected, we use \LProlog abstraction to capture their binding
structure.
Thus, $\letexp x n x$ is encoded as \lsti|(let (nat n) (x\x))|.
Similarly, the \LProlog term \lsti|(fix (f\x\ app f x))| represents
the source language expression $(\fix f x {f \app x})$.

We will use the \LProlog type \lsti|tm'| for encodings of target
language terms. To represent the constructs the target language shares
with the source language, we will use ``primed'' versions of the
\LProlog constants seen earlier; \eg, \lsti|unit'| of type \lsti|tm'|
will represent the null tuple. Of course, there will be no
constructor corresponding to \lsti|fix|. We will also
need the following additional constructors:

\vspace{-0.1cm}
{\footnotesize
\begin{tabbing}
\quad\=\kill
\>\lsti|abs'|\ :\ \lsti|(tm' -> tm') -> tm'|\qquad
\lsti|clos'|\ :\ \lsti|tm' -> tm' -> tm'|\\
\>\lsti|open'|\ :\ \lsti|tm' -> (tm' -> tm' -> tm') -> tm'|
\end{tabbing}
}
\vspace{-0.1cm}
\noindent Here, \lsti|abs'| encodes $\lambda$-abstraction and
\lsti|clos'| and \lsti|open'| encode closures and their
application.
Note again the $\lambda$-tree syntax representation for
binding constructs.

Following Section~\ref{sec:framework}, we represent typing judgments
as relations between terms and types, treating contexts implicitly via
dynamically added clauses that assign types to free variables.
We use the predicates \lsti|of| and \lsti|of'| to encode typing in the
source and target language respectively.
The clauses defining these predicates are routine and we show only a
few pertaining to the binding constructs.
The rule for typing fixed points in the source language translates
into the following.
\vspace{-0.1cm}
\begin{lstlisting}
of (fix R) (arr T1 T2) :- 
   pi f\ pi x\ of f (arr T1 T2) => of x T1 => of (R f x) T2.
\end{lstlisting}
\vspace{-0.1cm}
Note how the required freshness constraint is realized in this
clause: the universal quantifiers over \lsti|f| and
\lsti|x| introduce new names and the application
\lsti|(R f x)| replaces the bound variables with these names to
generate the new typing judgment that must be derived.
For the target language, the main interesting rule is for typing
the application of closures.
The following clause encodes this rule.
\vspace{-0.1cm}
\begin{lstlisting}
of' (open' M R) T :-  of' M (arr T1 T2),
    pi$\;$f\$\;$pi$\;$e\$\;$pi l\ of' f (arr' (prod (arr T1 T2) (prod T1 l)) T2) =>
                          of' e l => of' (R f e) T.
\end{lstlisting}
\vspace{-0.1cm}
Here again we use universal quantifiers in goals to encode
the freshness constraint.
Note also how the universal quantifier over the variable \lsti|l|
captures the opaqueness quality of the
type of the environment of the closure involved in the construct.

We encode the one step evaluation rules for the source and target
languages using the predicates \lsti|step| and
\lsti|step'|. We again consider
only a few interesting cases in their definition. Assuming that
\lsti|val| and \lsti|val'| recognize values in the
source and target languages, the clauses for evaluating the
application of a fixed point and a closure are the following.
\vspace{-0.1cm}
{\footnotesize
\begin{lstlisting}
step (app (fix R) V) (R (fix R) V) :- val V.
step' (open' (clos' F E) R) (R F E) :- val' (clos' F E).
\end{lstlisting}
}
\vspace{-0.1cm}
\noindent Note here how application in the
meta-language realizes substitution.

We use the predicates \lsti|nstep| (which relates a natural number and
two terms) and \lsti|eval| to represent the $n$-step and full
evaluation relations for the source language, respectively.
These predicates have obvious definitions.
The predicates \lsti|nstep'| and \lsti|eval'| play a similar role for
the target language.

\vspace{-0.4cm}
\subsubsection{Specifying closure conversion.}
\label{sssec:encode_ccrules}

To define closure conversion in \LProlog, we need a representation of
mappings for source language variables.
We use the type \lsti|map| and the constant
\lsti|map : tm -> tm' -> map| to represent the mapping for a single
variable.\footnote{This
  mapping is different from the one considered in
  Section~\ref{sec:twolevel} in that it is from a \emph{source} language
  variable to a \emph{target} language term.}
We use the type \lsti|map_list| for lists of such mappings, the
constructors \lsti|nil| and \lsti|::| for constructing such
lists and the predicate \lsti|member| for checking membership in them.
We also need to represent lists of source and target language terms.
We will use the types \lsti|tm_list| and \lsti|tm'_list| for these and
for simplicity of discussion, we will overload the list constructors
and predicates at these types.
Polymorphic typing in \LProlog supports such overloading but this
feature has not yet been implemented in Abella; we overcome this
difficulty in the actual development by using different type
and constant names for each case.

The crux in formalizing the definition of closure conversion is
capturing the content of the $\ccfix$ rule.
A key part of this rule is identifying the free variables in a given
source language term.
We realize the requirement by defining a predicate \lsti|fvars| that
is such that
if \lsti|(fvars M L1 L2)| holds then \lsti|L1| is a list that includes
all the free variables of \lsti|M| and \lsti|L2| is another list
that contains only the free variables of \lsti|M|.
We show a few critical clauses in the definition of this predicate,
omitting ones whose structure is easy predict.
\vspace{-0.1cm}
\begin{lstlisting}
fvars X _ nil :- notfree X.
fvars Y Vs (Y :: nil) :- member Y Vs.
fvars (nat _) _ nil.
fvars (plus M1 M2) Vs FVs :-
  fvars M1 Vs FVs1, fvars M2 Vs FVs2, combine FVs1 FVs2 FVs.
...
fvars (let M R) Vs FVs :- fvars M Vs FVs1,
  (pi x\ notfree x => fvars (R x) Vs FVs2), combine FVs1 FVs2 FVs.
fvars (fix R) Vs FVs :-
  pi f\ pi x\ notfree f => notfree x => fvars (R f x) Vs FVs.
\end{lstlisting}
\vspace{-0.1cm}
The predicate \lsti|combine| used in these clauses is one that holds
between three lists when the last is a combination of the elements of
the first two.
The essence of the definition of \lsti|fvars| is in the treatment of binding
constructs.
Viewed operationally, the body of such a construct is
descended into after instantiating the binder with a new variable
marked \lsti|notfree|.
Thus, the variables that are marked in this way
correspond to exactly those that are explicitly bound in the term and
only those that are not so marked are collected through the second
clause.
It is important also to note that the specification of \lsti|fvars|
has a completely logical structure; this fact can be exploited during
verification.

The $\ccfix$ rule requires us to construct an environment representing
the mappings for the variables found by \lsti|fvars|.
The predicate
\lsti|mapenv|
specified by the following clauses provides this functionality.
\vspace{-0.1cm}
\begin{lstlisting}
mapenv nil _ unit.
mapenv (X::L) Map (pair' M ML) :- member$\;$(map X M)$\;$Map, mapenv L Map ML.
\end{lstlisting}
\vspace{-0.1cm}
\noindent The $\ccfix$ rule also requires us to create a new mapping
from the variable list to projections from an environment variable.
Representing the list of projection mappings as a function from the
environment variable, this relation is given by the predicate
\lsti|mapvar| that is defined by the following clauses.
\vspace{-0.1cm}
\begin{lstlisting}
mapvar nil (e\ nil).
mapvar (X::L) (e\ (map X (fst' e))::(Map (snd' e))) :- mapvar L Map.
\end{lstlisting}
\vspace{-0.1cm}

We can now specify the closure conversion
transformation. We provide clauses below that define the
predicate \lsti|cc| such that \lsti|(cc Map Vs M M')| holds if
\lsti|M'| is a transformed
version of \lsti|M| under the mapping \lsti|Map| for the variables in
\lsti|Vs|; we assume that \lsti|Vs| contains all the free
variables of \lsti|M|.
\begin{lstlisting}
cc _ _ (nat N) (nat' N).
cc Map Vs X M :- member (map X M) Map.
cc Map Vs (pred M) (pred' M') :-  cc Map Vs M M'.
cc Map Vs (plus M1 M2) (plus' M1' M2') :-
   cc Map Vs M1 M1', cc Map Vs M2 M2'.
cc Map Vs (ifz M M1 M2) (ifz' M' M1' M2') :-
   cc Map Vs M M', cc Map Vs M1 M1', cc Map Vs M2 M2'.
cc Map Vs unit unit'.
cc Map Vs (pair M1 M2) (pair' M1' M2') :-
   cc Map Vs M1 M1', cc Map Vs M2 M2'.
cc Map Vs (fst M) (fst' M') :- cc Map Vs M M'.
cc Map Vs (snd M) (snd' M') :- cc Map Vs M M'.
cc Map Vs (let M R) (let' M' R') :- cc Map Vs M M',
  pi x\ pi y\ cc ((map x y) :: Map) (x :: Vs) (R x) (R' y).
cc Map Vs (fix R) (clos' (abs' (p\ let' (fst' p) (g\
                      let' (fst' (snd' p)) (y\
                      let' (snd' (snd' p)) (e\ R' g y e))))) E) :-
  fvars (fix R) Vs FVs, mapenv FVs Map E, mapvar FVs NMap,
  pi f\ pi x\ pi g\ pi y\ pi e\
   cc ((map x y)::(map f g)::(NMap e)) (x::f::FVs) (R f x)  (R' g y e).
cc Map Vs (app' M1 M2)
   (let' M1' (g\ open' g (f\e\ app' f (pair' g (pair' M2' e))))) :-
  cc Map Vs M1 M1', cc Map Vs M2 M2'.
\end{lstlisting}
These clauses correspond very closely to the rules in
Figure~\ref{fig:cc}.
Note especially the clause for transforming an expression of the form
\lsti|(fix R)| that encodes the content of the $\ccfix$ rule.
In the body of this clause, \lsti|fvars| is used to identify the free
variables of the expression, and \lsti|mapenv| and \lsti|mapvar| are
used to create the reified environment and the new mapping.
In both this clause and in the one for transforming a \lsti|let|
expression, the $\lambda$-tree representation, universal goals and
(meta-language) applications are used to encode freshness and renaming
requirements related to bound variables in a concise and logically
precise way.

\vspace{-0.3cm}
\subsection{Implementing other transformations}
\label{ssec:implothers}

We have used the ideas discussed in the preceding subsections in
realizing other transformations such as code hoisting and conversion
to continuation-passing style (CPS).
These transformations are part of a tool-kit used by
compilers for functional languages to convert programs into a form
from which compilation may proceed in a manner similar to that for
conventional languages like C.

Our implementation of the CPS transformation is based on the one-pass
version described by Danvy and Filinski~\cite{danvy92mscs} that
identifies and eliminates the so-called administrative redexes
on-the-fly.
This transformation can be encoded concisely and elegantly in \LProlog
by using meta-level redexes for administrative redexes.
The implementation is straightforward and similar ones that use the
\HOAS approach have already been described in the literature;~\eg
see~\cite{tian06cats}.

Our implementation of code hoisting is more interesting: it benefits
in an essential way once again from the ability to analyze binding
structure.
The code hoisting transformation lifts nested functions that
are closed out into a flat space at the top level in the program.
This transformation can be realized as a recursive procedure: given a
function $(\abs x M)$, the procedure is applied to the subterms of $M$
and the extracted functions are then moved out of $(\abs x M)$.
Of course, for this movement to be possible, it must be the case that
the variable $x$ does not appear in the functions that are candidates
for extraction.
This ``dependency checking'' is easy to encode in a logical way
within our framework.

To provide more insight into our implementation of code-hoisting, let
us assume that it is applied after closure conversion and that its
source and target languages are both the language shown in
Figure~\ref{fig:targlang}.
Applying code hoisting to any term will result in a term of the form
\vspace{-0.1cm}
\begin{smallgather}
\letexp {f_1} {M_1} {\ldots \letexp {f_n} {M_n} M}
\end{smallgather}

\vspace{-0.1cm}
\noindent where, for $1 \leq i \leq n$, $M_i$ corresponds to an
extracted function. 
We will write this term below as $(\letfun {\vctr{f} = \vctr{M}} M)$
where $\vctr{f} = (f_1,\ldots,f_n)$ and, correspondingly, $\vctr{M} =
(M_1,\ldots,M_n)$. 

We write the judgment of code hoisting as $(\ch \rho M {M'})$ where
$\rho$ has the form $(x_1,\ldots,x_n)$.
This judgment asserts that $M'$ is the result of extracting all
functions in $M$ to the top level, assuming that $\rho$ contains all
the bound variables in the context in which $M$ appears.
The relation is defined by recursion on the structure of $M$.
The main rule that deserves discussion is that for
transforming functions. This rule is the following:

\vspace{-0.1cm}
\begin{smallgather}
    \infer{
      \ch \rho {\abs x M}
          {\letfun {(\vctr{f},g) = (\vctr{F},\abs {f} {\abs x {\letfun
                  {\vctr{f} = (\pi_1(f),\ldots,\pi_n(f))} {M'}}})}
            {g~\vctr{f}}}
    }{
      \ch {\rho,x} M {\letfun {\vctr{f} = \vctr{F}} {M'}}
    }
\end{smallgather}

\vspace{-0.1cm}
We assume here that $\vctr{f}=(f_1,\ldots,f_n)$ and, by an abuse of
notation, we let $(g~\vctr{f})$ denote $(g~(f_1,\ldots,f_n))$.
This rule has a side condition: $x$ must not occur in $\vctr{F}$.
Intuitively, the term $(\abs x M)$, is transformed by extracting the
functions from within $M$ and then moving them further out of
the scope of $x$.
Note that this transformation succeeds only if none of the
extracted functions depend on $x$.
The resulting function is then itself extracted. In order to do this,
it must be made independent of the (previously) extracted functions,
something that is achieved by a suitable abstraction; the expression
itself becomes an application to a tuple of functions in an
appropriate let environment.

It is convenient to use a special representation for the result of
code hoisting in specifying it in \LProlog.
Towards this end, we introduce the following constants:

\vspace{-0.1cm}
{\footnotesize
\begin{tabbing}
\quad\=\kill
\>\lsti+hbase : tm' -> tm'+\\
\>\lsti+habs : (tm' -> tm') -> tm'+\\
\>\lsti+htm : tm'_list -> tm' -> tm'+
\end{tabbing}
}

\vspace{-0.1cm}
\noindent Using these constants, the term
$(\letfun {(f_1,\ldots,f_n) = (M_1,\ldots,M_n)} M)$ that results from
code hoisting will be represented by

\vspace{-0.1cm}
\begin{lstlisting}
  htm (M1 :: ... :: Mn :: nil) (habs (f1\ ... (habs (fn\ hbase M)))).
\end{lstlisting}

\vspace{-0.1cm}
\noindent We use the predicate \lsti+ch : tm' -> tm' -> o+ to represent the
code hoisting judgment.
The context $\rho$ in the judgment will be encoded implicitly through
dynamically added program clauses that specify the translation of each
variable \lsti+x+ as \lsti+(htm nil (hbase x))+.
In this context, the rule for transforming functions, the main rule
of interest, is encoded in the following clause:
\vspace{-0.1cm}
\begin{lstlisting}
ch (abs' M) M'' :-
   (pi x\ ch x (htm nil (hbase x)) => ch (M x) (htm FE (M' x))),
   extract FE M' M''.
\end{lstlisting}
\vspace{-0.1cm}
\noindent As in previous specifications, a universal and a
   hypothetical goal are used in this clause to realize recursion over
   binding structure. 
Note also the completely logical encoding of the requirement that
the function argument must not occur in the nested functions extracted
from its body: quantifier ordering ensures that \lsti+FE+
cannot be instantiated by a term that contains \lsti|x| free in it.
We have used the predicate \lsti|extract| 
to build the final result of the transformation from the transformed
form of the function body and the nested functions extracted from it;
the definition of this predicate is easy to construct and is not
provided here.

\section{Verifying Transformations on Functional Programs}
\label{sec:verifpt}

We now consider the verification of \LProlog implementations of
transformations on functional programs. We exploit the two-level logic
approach in this process, treating \LProlog programs as \HOHH
specifications and reasoning about them using Abella. Our discussions
below will show how we can use the $\lambda$-tree syntax approach and
the logical nature of our specifications to benefit in the reasoning
process. Another aspect that they will bring out is the virtues of the
close correspondence between rule based presentations and \HOHH
specifications: this correspondence allows the structure of informal
proofs over inference rule style descriptions to be mimicked in a
formalization within our framework.

We use the closure conversion transformation as our main
example in this exposition. The first two subsections below
present, respectively, an informal proof of its correctness and its
rendition in Abella. We then discuss the application of these ideas
to other transformations. Our proofs are based on logical relation
style definitions of program equivalence. Other forms of semantics
preservation have also been considered in the literature. Our
framework can be used to advantage in formalizing these approaches as
well, an aspect we discuss in the last subsection.

\vspace{-0.3cm}
\subsection{Informal verification of closure conversion}
\label{ssec:ccproof}

To prove the correctness of closure conversion, we need a notion
of equivalence between the source and target programs.
Following~\cite{minamide95tr}, we use a logical relation style
definition for this purpose.
A complication is that our source language includes recursion.
To overcome this problem, we use the idea of step
indexing~\cite{ahmed06esop,appel01toplas}.
Specifically, we define the following mutually recursive
simulation relation $\sim$ between
closed source and target terms and equivalence relation $\approx$
between closed source and target values, each indexed by a
type and a step measure.

\begin{smallalign}
  & \simulate T k M M' \iff \forall j \leq k. \forall V. M \step{j} V \imply
      \exists V'. {M'} \eval {V'} \conj \equal T {k-j} V {V'};\\
  & \equal \tnat k n n; \qquad
    \equal \tunit k \unit \unit;\\
  & \equal {(T_1 \tprod T_2)} k {\pair {V_1} {V_2}} {\pair {V_1'} {V_2'}} \iff
       \equal {T_1} k {V_1} {V_1'} \conj \equal {T_2} k {V_2} {V_2'};\\
  & \equal {T_1 \to T_2} k
           {(\fix f x M)}
           {\clos {V'} {V_e}} \iff  \forall j < k. \forall V_1, V_1', V_2, V_2'.\\
  & \qquad \equal {T_1} j {V_1} {V_1'} \imply
           \equal {T_1 \to T_2} j {V_2} {V_2'} \imply
           \simulate {T_2} j {M[V_2/f, V_1/x]} {V' \app (V_2', V_1', V_e)}.
\end{smallalign}
Note that the definition of $\approx$ in the fixed point/closure case
uses $\approx$ negatively at the same type. However, it is still a
well-defined notion because the index decreases. The cumulative notion
of equivalence, written $\simulatesans{T}{M}{M'}$, corresponds to two
expressions being equivalent under \emph{any} index.

Analyzing the simulation relation and using the evaluation rules,
we can show the following ``compatibility'' lemma for various
constructs in the source language.
\begin{mylemma}\label{lem:sim_compose}

\begin{enumerate}
  \item If $\simulate \tnat k {M} {M'}$ then $\simulate \tnat k {\pred M}
  {\pred M'}$. If also $\simulate \tnat k {N} {N'}$ then $\simulate
  \tnat k {M + N} {M' + N'}$.

  \item  If $\simulate {T_1 \tprod T_2} k {M} {M'}$ then $\simulate {T_1} k
        {\fst{M}} {\fst{M'}}$ and $\simulate {T_2} k {\snd{M}}
        {\snd{M'}}$.

  \item If $\simulate {T_1} {k} {M} {M'}$ and $\simulate {T_2} {k} {N}
        {N'}$ then $\simulate {T_1 \tprod T_2} {k} {(M,N)} {(M',N')}.$

  \item If $\simulate \tnat k M {M'}$, $\simulate T k {M_1} {M_1'}$ and
      $\simulate T k {M_2} {M_2'}$, then\\
      $\simulate T k {\ifz M {M_1} {M_2}} {\ifz {M'} {M_1'} {M_2'}}$.

  \item If $\simulate{T_1 \to T_2}{k}{M_1}{M_1'}$ and
  $\simulate{T_1}{k}{M_2}{M_2'}$ then\\
  $\simulate{T_2}{k} {M_1 \app M_2} {\letexp g {M_1'}
     {\open {x_f} {x_e} {g}{x_f \app (g,M_2',x_e)}}}.$
\end{enumerate}
\end{mylemma}
\noindent The proof of the last of these properties requires us to
consider the evaluation of the application of a fixed point expression
which involves ``feeding'' the expression to its own body. In working
out the details, we use the easily observed property that the
simulation and equivalence relations are closed under decreasing
indices.

Our notion of equivalence only relates closed terms.
However, our transformation typically operates on
open terms, albeit under mappings for the free variables.
To handle this situation, we consider semantics preservation for
possibly open terms under closed substitutions.
We will take substitutions in both the source and
target settings to be simultaneous mappings of closed values for a
finite collection of variables, written as
$(V_1/x_1,\ldots,V_n/x_n)$.
In defining a correspondence between source and target language
substitutions, we need to consider the possibility that a collection
of free variables in the first may be reified into an environment
variable in the second.
This motivates the following definition in which $\gamma$ represents a
source language substitution:
\begin{smallalign}
  & \equal {x_m:T_m, \ldots, x_1:T_1}
           k
           {\gamma}
           {(V_1,\ldots,V_m)}
    \iff
    \forall 1 \leq i \leq m. \equal {T_i} k {\gamma(x_i)} {V_i}.
\end{smallalign}
Writing $\substconcat{\gamma_1}{\gamma_2}$ for the concatenation of
two substitutions viewed as lists, equivalence between substitutions
is then defined as follows:
\begin{smallalign}
 &    \equal {\Gamma, x_n:T_n, \ldots,x_1:T_1}
           k
           {\substconcat{(V_1/x_1,\ldots,V_n/x_n) }{\gamma}}
           {(V_1'/y_1, \ldots, V_n'/y_n, V_e/x_e)}\\
  & \qquad  \iff
    (\forall 1 \leq i \leq n. \equal {T_i} k {V_i} {V_i'}) \conj
    \equal \Gamma k \gamma {V_e}.
\end{smallalign}
Note that both relations are indexed by a source language
typing context and a step measure. The second relation allows the
substitutions to be for different variables in the source 
and target languages. A relevant mapping will determine a
correspondence between these variables when we use the relation.

We write the application of a substitution $\gamma$ to a term $M$ as
$M[\gamma]$.
The first part of the following lemma, proved by an easy use of the definitions of
$\approx$ and evaluation, provides the basis for justifying the
treatment of free variables via their transformation into projections
over environment variables introduced at function boundaries in the
closure conversion transformation. The second part of the lemma is a
corollary of the first part that relates a source substitution
and an environment computed during the closure conversion of fixed points.
\begin{mylemma}\label{lem:var_sem_pres}
  Let $\delta = \substconcat{(V_1/x_1,\ldots,V_n/x_n)}{\gamma}$ and
  $\delta' = (V_1'/y_1,\ldots,$ $V_n'/y_n,V_e/x_e)$
  be source and target language substitutions and let $\Gamma =
  (x_m':T_m',\ldots,x_1':T_1',x_n:T_n,\ldots,x_1:T_1)$ be a source language
  typing context such that $\equal \Gamma k \delta
  {\delta'}$. Further, let $\rho = (x_1  \mapsto y_1,\ldots,x_n \mapsto
  y_n, x_1' \mapsto \pi_1(x_e), \ldots, x_m' \mapsto \pi_m(x_e))$.
\begin{enumerate}
\item If $x : T \in \Gamma$ then there exists a value $V'$ such that
  $(\rho(x))[\delta'] \eval V'$ and $\equal T k {\delta(x)} {V'}$.

\item   If $\Gamma' = (z_1:T_{z_1},\ldots,z_j:T_{z_j})$ for $\Gamma'
  \subseteq \Gamma$ and $\ccenv \rho {(z_1,\ldots,z_j)} M$, then
  there exists $V_e'$ such that $M[\delta'] \eval V_e'$ and $\equal {\Gamma'} k
  \delta {V_e'}$.
\end{enumerate}
\end{mylemma}
The proof of semantics preservation also requires a result about the
preservation of typing.
It takes a little effort to ensure that this property holds at the
point in the transformation where we cross a function boundary.
That effort is encapsulated in the following strengthening lemma 
in the present setting.
\begin{mylemma}\label{lem:typ_str}
  If $\Gamma \tseq M:T$, $\{x_1,\ldots,x_n\} \supseteq \fvars M$ and
  $x_i:T_i \in \Gamma$ for $1 \leq i \leq n$, then $x_n:T_n
  ,\ldots,x_1:T_1 \tseq M :T$.
\end{mylemma}

The correctness theorem can now be stated as follows:
\begin{mythm}\label{thm:cc_sem_pres}
  Let $\delta = \substconcat{(V_1/x_1,\ldots,V_n/x_n)}{\gamma}$ and
  $\delta' = (V_1'/y_1,\ldots,$ $V_n'/y_n,V_e/x_e)$
  be source and target language substitutions and let $\Gamma =
  (x_m':T_m',\ldots,x_1':T_1',x_n:T_n,\ldots,x_1:T_1)$ be a source language
  typing context such that $\equal \Gamma k \delta
  {\delta'}$. Further, let $\rho = (x_1  \mapsto y_1,\ldots,x_n \mapsto
  y_n, x_1' \mapsto \pi_1(x_e), \ldots, x_m' \mapsto \pi_m(x_e))$.
  If $\Gamma \tseq M:T$ and $\cc \rho M M'$, then $\simulate T k
  {M[\delta]} {M'[\delta']}$.
\end{mythm}
\noindent We outline the main steps in the argument for this theorem:
these will guide the development of a formal proof in
Section~\ref{ssec:vericc}.
We proceed by induction on the derivation of $\cc \rho M M'$,
analyzing the last step in it.
This obviously depends on the structure of $M$.
The case for a number is obvious and for a variable we
use Lemma~\ref{lem:var_sem_pres}.1.
In the remaining cases,  other than when $M$ is of the form $(\letexp x
{M_1} {M_2})$ or $(\fix f x {M_1})$, the argument follows a set pattern: we
observe that substitutions distribute to the sub-components of
expressions, we invoke the induction hypothesis over the
sub-components and then we use Lemma~\ref{lem:sim_compose} to conclude.
If $M$ is of the form $(\letexp x {M_1} {M_2})$, then $M'$ must be of
the form $(\letexp y {M_1'} {M_2'})$. Here again the substitutions
distribute to $M_1$ and $M_2$ and to $M_1'$ and $M_2'$,
respectively. We then apply the induction hypothesis first to $M_1$
and $M_1'$ and then to $M_2$ and $M_2'$; in the latter
case, we need to consider extended substitutions but these
obviously remain equivalent.
Finally, if $M$ is of the form $(\fix f x {M_1})$, then $M'$ must have the
form $\clos {M_1'} {M_2'}$.
We can prove that the abstraction $M_1'$ is closed and therefore
that $M'[\sigma'] = \clos {M_1'} {M_2'[\sigma']}$.
We then apply the induction hypothesis.
In order to do so, we generate the appropriate typing judgment using
Lemma~\ref{lem:typ_str} and a new pair of equivalent substitutions
(under a suitable step index) using Lemma~\ref{lem:var_sem_pres}.2.

\vspace{-0.3cm}
\subsection{Formal verification of the implementation of closure conversion}
\label{ssec:vericc}

In the subsections below, we present a sequence of preparatory steps,
leading eventually to a formal version of the correctness theorem.

\vspace{-0.3cm}
\subsubsection{Auxiliary predicates used in the formalization.}
We use the techniques of Section~\ref{sec:framework} to define some
predicates related to the encodings of source and target language
types and terms that are needed in the main development; unless
explicitly mentioned, these definitions are in \Gee.
First, we define the predicates \lsti|ctx| and \lsti|ctx'| to identify
typing contexts for the source and target languages.
Next, we define in \HOHH the recognizers \lsti|tm| and \lsti|tm'| of
well-formed source and target language terms.
A source (target) term \lsti|M| is closed if \lsti|{tm M}|
(\lsti|{tm' M}|) is derivable.
The predicate \lsti|is_sty| recognizes source types. Finally,
\lsti|vars_of_ctx| is a predicate such that \lsti|(vars_of_ctx L Vs)|
holds if \lsti|L| is a source language typing context and \lsti|Vs|
is the list of variables it pertains to.

Step indexing uses ordering on natural numbers. We represent natural
numbers using \lsti|z| for 0 and \lsti|s| for the successor
constructor. The predicate \lsti|is_nat| recognizes natural numbers.
The predicates \lsti|lt| and \lsti|le|, whose definitions are routine,
represent the ``less than'' and the ``less than or equal to''
relations.

\vspace{-0.3cm}
\subsubsection{The simulation and equivalence relations.}
The following clauses define the simulation and equivalence
relations.
\begin{lstlisting}
sim T K M M' :=  forall J V, le J K -> {nstep J M V} ->  {val V} ->
  exists V' N, {eval' M' V'} /\ {add J N K} /\ equiv T N V V';
equiv tnat K (nat N) (nat' N);
equiv tunit K unit unit';
equiv (prod T1 T2) K (pair V1 V2) (pair' V1' V2') :=
  equiv T1 K V1 V1' /\ equiv T2 K V2 V2' /\
  {tm V1} /\ {tm V2} /\ {tm' V1'} /\ {tm' V2'};
equiv (arr T1 T2) z (fix R) (clos' (abs' R') VE) :=
  {val' VE} /\ {tm (fix R)} /\ {tm' (clos' (abs' R') VE)};
equiv (arr T1 T2) (s K) (fix R) (clos' (abs' R') VE)  :=
  equiv (arr T1 T2) K (fix R) (clos' (abs' R') VE) /\
  forall V1 V1' V2 V2', equiv T1 K V1 V1' -> equiv (arr T1 T2) K V2 V2' ->
     sim T2 K (R V2 V1) (R' (pair' V2' (pair' V1' VE))).
\end{lstlisting}
The formula \lsti|(sim T K M M')| is intended to mean that
\lsti|M| simulates \lsti|M'| at type \lsti|T| in \lsti|K| steps;
\lsti|(equiv T K V V')| has a similar interpretation.
Note the exploitation of $\lambda$-tree syntax, specifically the use
of application, to realize substitution in the definition of
\lsti|equiv|.
It is easily shown that \lsti|sim| holds only between closed source
and target terms and similarly \lsti|equiv| holds only between closed
source and target values.\footnote{The definition of \lsti|equiv|
uses itself negatively in the last clause and thereby
violates the original stratification condition of \Gee. However,
Abella permits this definition under a weaker stratification condition
that ensures consistency provided the definition is used in 
restricted ways~\cite{baelde12lics,tiu12ijcar}, a requirement that is
adhered to in this paper.} 

Compatibility lemmas in the style of Lemma \ref{lem:sim_compose} are
easily stated for \lsti|sim|. For example, the one for pairs is the following.
\begin{lstlisting}
forall T1 T2 K M1 M2 M1' M2', {is_nat K} -> {is_sty T1} -> {is_sty T2} ->
  sim T1 K M1 M1' -> sim T2 K M2 M2' ->
  sim (prod T1 T2) K (pair M1 M2) (pair' M1' M2').
\end{lstlisting}
These lemmas have straightforward proofs.

\vspace{-0.3cm}
\subsubsection{Representing substitutions.}\label{sec:expl_subst}
We treat substitutions as discussed in
Section~\ref{sec:framework}. For example, source substitutions satisfy
the following definition.
\begin{lstlisting}
subst nil;
subst ((map X V)::ML) := subst ML /\ name X /\ {val V} /\ {tm V} /\
  forall V', member (map X V') ML -> V' = V.
\end{lstlisting}
By definition, these substitutions map variables to closed
values. To accord with the way closure conversion is formalized, we
allow multiple mappings for a given variable, but we require all of
them to be to the same value.
The application of a source substitution is also defined as discussed
in Section~\ref{sec:framework}.
\begin{lstlisting}
app_subst nil M M;
nabla x,app_subst$\;$((map x V)::(ML x))$\;$(R x)$\;$M := nabla x,app_subst (ML x) (R V) M.
\end{lstlisting}
As before, we can easily prove properties about substitution
application based on this definition such as that such an application
distributes over term structure and that closed terms are not affected by
substitution.

The predicates \lsti|subst'| and \lsti|app_subst'| encode target
substitutions and their application. Their formalization is similar to
that above.

\vspace{-0.3cm}
\subsubsection{The equivalence relation on substitutions.}
We first define the relation \lsti|subst_env_equiv|
between source substitutions and target environments:
\begin{lstlisting}
  subst_env_equiv nil K ML unit';
  subst_env_equiv ((of X T)::L) K ML (pair' V' VE) :=
    exists V,subst_env_equiv$\;$L K ML VE /\ member$\;$(map X V)$\;$ML /\ equiv$\;$T K V V'.
\end{lstlisting}
Using \lsti|subst_env_equiv|, the needed relation between source
and target substitutions is defined as follows.
\begin{lstlisting}
nabla e, subst_equiv L K ML ((map e VE)::nil) := subst_env_equiv L K ML VE;
nabla x y, subst_equiv ((of x T)::L) K ((map x V)::ML) ((map y V')::ML') :=
  equiv T K V V' /\ subst_equiv L K ML ML'.
\end{lstlisting}

\vspace{-0.3cm}
\subsubsection{Lemmas about \lsti|fvars|, \lsti|mapvar| and \lsti|mapenv|.}

Lemma~\ref{lem:typ_str} translates into a lemma about \lsti|fvars| in
the implementation. To state it, we define a strengthening relation
between source typing contexts:
\begin{lstlisting}
prune_ctx$\;$nil L nil;
prune_ctx$\;$(X::Vs)$\;$L$\;$((of$\;$X$\;$T)::L') := member$\;$(of$\;$X$\;$T)$\;$L /\ prune_ctx$\;$Vs$\;$L$\;$L'.
\end{lstlisting}
\lsti|(prune_ctx Vs L L')| holds if \lsti|L'| is a typing context
that ``strengthens'' \lsti|L| to contain type assignments only for the
variables in \lsti|Vs|.
The lemma about \lsti|fvars| is then the following.
\begin{lstlisting}
forall L Vs M T FVs, ctx L -> vars_of_ctx L Vs -> {L |- of M T} ->
  {fvars M Vs FVs} -> exists L', prune_ctx FVs L L' /\ {L' |- of M T}.
\end{lstlisting}
To prove this theorem, we generalize it so that the \HOHH
derivation of \lsti|(fvars M Vs FVs)| is relativized to a context
that marks some variables as not free. The resulting generalization is
proved by induction on the \lsti|fvars| derivation.

A formalization of Lemma \ref{lem:var_sem_pres} is also needed for the
main theorem. We start with a lemma about \lsti|mapvar|.
\begin{lstlisting}
forall L Vs Map ML K VE X T M' V, nabla e, {is_nat K} -> ctx L -> subst ML ->
  subst_env_equiv L K ML VE -> vars_of_ctx L Vs -> {mapvar Vs Map} ->
  member (of X T) L ->   app_subst ML X V -> {member$\;$(map$\;$X$\;$(M'$\;$e))$\;$(Map$\;$e)}
  -> exists V', {eval' (M' VE) V'} /\ equiv T K V V'.
\end{lstlisting}
In words, this lemma states the following. If \lsti|L| is a source
typing context for the variables $(x_1,\ldots,x_n)$, \lsti|ML| is a
source substitution and \lsti|VE| is an environment equivalent to
\lsti|ML| at \lsti|L|, then \lsti|mapvar| determines a mapping for
$(x_1,\ldots,x_n)$ that are projections over an environment with the
following character: if the environment is taken to be \lsti|VE|,
then, for $1 \leq i \leq n$, $x_i$ is mapped to a projection that must
evaluate to a value equivalent to the substitution for $x_i$ in
\lsti|ML|. The lemma is proved by induction on the derivation of
\lsti|{mapvar Vs Map}|.

Lemma \ref{lem:var_sem_pres} is now formalized as follows.
\begin{lstlisting}
forall L ML ML' K Vs Vs' Map, {is_nat K} -> ctx L -> subst ML ->
  subst' ML' -> subst_equiv L K ML ML' -> vars_of_ctx L Vs ->
  vars_of_subst' ML' Vs' -> to_mapping Vs Vs' Map ->
  (forall X T V M' M'', member (of X T) L -> {member (map X M') Map} ->
    app_subst ML X V -> app_subst' ML' M' M'' ->
      exists V', {eval' M'' V'} /\ equiv T K V V') /\
  (forall L' NFVs E E', prune_ctx NFVs L L' ->
    {mapenv NFVs Map E} -> app_subst' ML' E E' ->
      exists VE', {eval' E' VE'} /\ subst_env_equiv L' K ML VE').
\end{lstlisting}
Two new predicates are used here. The judgment
\lsti|(vars_of_subst' ML' Vs')|  ``collects'' the variables in the target
substitution \lsti|ML'| into \lsti|Vs'|. Given
source variables \lsti|Vs = $(x_1,\ldots,x_n,x_1',\ldots,x_m')$|
and target variables \lsti|Vs' = $(y_1,\ldots,y_n,x_e)$|,
the predicate \lsti|to_mapping| creates in \lsti|Map| the  mapping

\vspace{-0.2cm}
\begin{tabbing}
\qquad\=\kill
\>$(x_1  \mapsto y_1,\ldots,x_n \mapsto y_n,
x_1' \mapsto \pi_1(x_e), \ldots, x_m' \mapsto \pi_m(x_e))$.
\end{tabbing}

\vspace{-0.2cm}
\noindent The conclusion of the lemma is a conjunction representing
the two parts of Lemma \ref{lem:var_sem_pres}. The first part
is proved by induction on \lsti|{member (map X M') Map}|, using
the lemma for \lsti|mapvar| when \lsti|X| is some $x_i' (1 \leq i \leq
m)$. The second part is proved by induction on
\lsti|{mapenv NFVs Map E}| using the first part.

\vspace{-0.3cm}
\subsubsection{The main theorem.}
The semantics preservation theorem is stated as follows:
\begin{lstlisting}
forall L ML ML' K Vs Vs' Map T P P' M M', {is_nat K} -> ctx L -> subst ML ->
  subst' ML' -> subst_equiv L K ML ML' -> vars_of_ctx L Vs ->
  vars_of_subst' ML' Vs' -> to_mapping Vs Vs' Map -> {L |- of M T} ->
  {cc$\;$Map$\;$Vs$\;$M$\;$M'} -> app_subst$\;$ML$\;$M$\;$P -> app_subst'$\;$ML'$\;$M'$\;$P' -> sim$\;$T$\;$K$\;$P$\;$P'.
\end{lstlisting}
We use an induction on \lsti|{cc Map Vs M M'}|, the closure conversion
derivation, to prove this theorem. As should
be evident from the preceding development, the proof in fact closely
follows the structure we outlined in Section~\ref{ssec:ccproof}.

\vspace{-0.3cm}
\subsection{Verifying the implementations of other transformations}
\label{ssec:veriothers}

We have used the ideas presented in this section to develop semantics
preservation proofs for other transformations such as code hoisting
and the CPS transformation. We discuss the case for code hoisting
below.

The first step is to define the step-indexed logical
relations $\chsim$ and $\approx'$ that respectively represent the
simulation and equivalence relation between the input and output terms
and values for code hoisting:
\begin{smallalign}
  & \chsimulate T k M M' \iff \forall j \leq k. \forall V. M \step{j} V \imply
      \exists V'. {M'} \eval {V'} \conj \chequal T {k-j} V {V'};\\[5pt]
  & \chequal \tnat k n n; \\
  & \chequal \tunit k \unit \unit;\\
  & \chequal {(T_1 \tprod T_2)} k {\pair {V_1} {V_2}} {\pair {V_1'} {V_2'}} \iff
       \chequal {T_1} k {V_1} {V_1'} \conj \chequal {T_2} k {V_2} {V_2'};\\
  & \chequal {T_1 \carr T_2} k {(\abs x M)} {(\abs x M')} \iff
       \forall j < k. \forall V, V'. \chequal {T_1} j V {V'} \imply
         \chsimulate {T_2} j {M[V/x]} {M'[V'/x]};\\
  & \chequal {T_1 \to T_2} k {\clos {\abs p M} {V_e}} {\clos {\abs p M'} {V_e'}} \iff
       \forall j < k. \forall V_1, V_1', V_2, V_2'. \\
  & \qquad
       \chequal {T_1} j {V_1} {V_1'} \imply
       \chequal {T_1 \to T_2} j {V_2} {V_2'} \imply
       \chsimulate {T_2} j {M[(V_2,V_1,V_e)/p]} {M'[(V_2',V_1',V_e')/p]}.
\end{smallalign}
We can show that $\chsim$ satisfies a set of compatibility
properties similar to Lemma~\ref{lem:sim_compose}.

We next define a step-indexed relation of equivalence between two
substitutions $\delta = {(V_1/x_1,\ldots,V_m/x_m)}$ and $\delta' =
{(V_1'/x_1,\ldots,V_m'/x_m)}$ relative to a typing context $\Gamma =
({x_m:T_m,\ldots, x_1:T_1})$:
\begin{smallalign}
  & \chequal \Gamma k \delta {\delta'}
    \iff
    \forall 1 \leq i \leq m. \chequal {T_i} k {V_i} {V_i'}.
\end{smallalign}

The semantics preservation theorem for code hoisting is stated as follows:
\begin{mythm}
  Let $\delta = {(V_1/x_1,\ldots,V_m/x_m)}$ and $\delta' =
  {(V_1'/x_1,\ldots,V_m'/x_m)}$ be substitutions for the language
  described in Figure~\ref{fig:targlang}. Let $\Gamma =
  (x_m:T_m,\ldots,x_1:T_1)$ be a typing context such that $\chequal
  \Gamma k \delta {\delta'}$. Further, let $\rho = (x_1,\ldots,x_m)$. If
  $\Gamma \tseq M:T$ and $\ch \rho M {M'}$ hold, then $\chsimulate T
  k {M[\delta]} {M'[\delta']}$ holds.
\end{mythm}
The theorem is proved by induction on the derivation for $\ch \rho M
{M'}$. The base cases follow easily, possibly using the fact that
$\chequal \Gamma k \delta {\delta'}$.
For the inductive cases, we observe that substitutions distribute to the
sub-components of expressions, we invoke the induction hypothesis over
the sub-components and we use the compatibility property of
$\chsim$. In the case of an abstraction, $\delta$ and $\delta'$ must be
extended to include a substitution for the bound variable. For this
case to work out, we must show that the additional substitution
for the bound variable has no impact on the functions extracted by
code hoisting. From the side condition for the rule deriving $\ch \rho M
{M'}$ in this case, the extracted functions cannot depend on the bound
variable and hence the desired observation follows.

In the formalization of this proof, we use the predicate constants
\lsti+sim'+ and \lsti+equiv'+ to respectively represent $\chsim$ and
$\approx'$. The Abella definitions of these predicates have by now
a familiar structure. We also define a constant \lsti+subst_equiv'+ to
represent the equivalence of substitutions as follows:
\begin{lstlisting}
  subst_equiv' nil K nil nil;
  nabla x, subst_equiv'$\;$((of' x T)::L) K ((map' x V)::ML)$\;$((map' x V')::ML')
    := equiv' T K V V' /\ subst_equiv' L K ML ML'.
\end{lstlisting}
The representation of contexts in the code hoisting judgment in the
\HOHH specification is captured by the predicate \lsti+ch_ctx+ that is
defined as follows:
\begin{lstlisting}
  ch_ctx nil;
  nabla x, ch_ctx (ch x (htm nil (hbase x)) :: L) := ch_ctx L.
\end{lstlisting}
The semantics preservation theorem is stated as follows, where
\lsti|vars_of_ctx'| is a predicate for collecting variables in the
typing contexts for the target language, \lsti|vars_of_ch_ctx| is a
predicate such that \lsti|(vars_of_ch_ctx L Vs)| holds if \lsti|L| is
a context for code hoisting and \lsti|Vs| is the list of variables it
pertains to:
\begin{lstlisting}
forall L K CL ML ML' M M' T FE FE' P P' Vs, {is_nat K} -> ctx' L ->
  ch_ctx CL -> vars_of_ctx' L Vs -> vars_of_ch_ctx CL Vs ->
  subst' ML -> subst' ML' ->  subst_equiv' L K ML ML' ->
  {L |- of' M T} -> {CL |- ch M (htm FE M')} -> app_subst' ML M P ->
  app_subst' ML' (htm FE M') (htm FE' P') -> sim' T K P (htm FE' P').
\end{lstlisting}
The proof is by induction on \lsti+{CL |-$\;$ch M (htm$\;$FE$\;$M')}+
and its structure follows that of the informal one very closely.
The fact that the extracted functions do not depend on the bound
variable of an abstraction is actually explicit in the logical
formulation and this leads to an exceedingly simple argument for this
case.

\vspace{-0.3cm}
\subsection{Relevance to other styles of correctness proofs}
\label{ssec:othercorrectness}

Many compiler verification projects, such as CompCert~\cite{leroy06popl} and
CakeML~\cite{kumar14popl}, have focused primarily on verifying whole programs
that produce values of atomic types.
In this setting, the main requirement is to show that the source and
target programs evaluate to the same atomic values.
Structuring a proof around program equivalence base on a logical
relation is one way to do this.
Another, sometimes simpler, approach is to show that the compiler
transformations permute over evaluation; this method works because
transformations typically preserve values at atomic types.
Although we do not present this here, we have examined proofs of this
kind and have observed many of the same kinds of benefits to
the $\lambda$-tree syntax approach in their context as well.

Programs are often built by composing separately compiled
modules of code.
In this context it is desirable that the composition of correctly
compiled modules preserve correctness; this property applied
to compiler verification has been called modularity.
Logical relations pay attention to equivalence at
function types and hence proofs based on them possess the modularity
property.
Another property that is desirable for correctness proofs is
transitivity: we should be able to infer the correctness of
a multi-stage compiler from the correctness of each of its stages.
This property holds when we use logical relations if we restrict
attention to programs that produce atomic values but cannot be
guaranteed if equivalence at function types is also important; it is
not always possible to decompose the natural logical relation between
a source and target language into ones between several intermediate
languages.
Recent work has attempted to generalize the logical relations based
approach to obtain the benefits of both transitivity
and modularity~\cite{neis15icfp}.
Many of the same issues relating to the treatment of binding and
substitution appear in this context as well and the work in
this paper therefore seems to be relevant also to the formalization of
proofs that use these ideas.

Finally, we note that the above comments relate only to the
formalization of proofs.
The underlying transformations remain unchanged and so does the
significance of our framework to their implementation.

\section{Related Work and Conclusion}
\label{sec:related}

Compiler verification has been an active area for investigation. 
We focus here on the work in this area that has been devoted to
compiling functional languages. 
There have been several projects with ambitious scope even in this
setting. 
To take some examples, the CakeML project has implemented a compiler
from a subset of ML to the X86 assembly language and verified it
using HOL4~\cite{kumar14popl};
Dargaye has used Coq to verify a compiler from a subset of ML into
the intermediate language used by CompCert~\cite{dargaye09phd};
Hur and Dreyer have used Coq to develop a verified single-pass
compiler from a subset of ML to assembly code based on a logical
relations style definition of program equivalence~\cite{hur11popl};
and Neis \etal\ have used Coq to develop a verified multi-pass
compiler called Pilsner, basing their proof on a notion of semantics
preservation called Parametric Inter-Languages Simulation
(PILS)~\cite{neis15icfp}. 
All these projects have used essentially first-order treatments of
binding, such as those based on a De Bruijn style representation.

A direct comparison of our work with the projects mentioned above is
neither feasible nor sensible because of differences in scope and
focus.  
Some comparison is possible with a part of the Lambda Tamer
project of Chlipala in which he describes the verified implementation
in Coq of a compiler for the STLC using a logical relation based
definition of program equivalence~\cite{chlipala07pldi}. 
This work uses a higher-order representation of syntax that does not
derive all the benefits of $\lambda$-tree syntax. 
Chlipala's implementation of closure conversion comprises about 400
lines of Coq code, in contrast to about 70 lines of \LProlog code that
are needed in our implementation. 
Chlipala's proof of correctness comprises about 270 lines but it
benefits significantly from the automation framework that was the
focus of the Lambda Tamer project; that framework is built on top of the
already existing Coq libraries and consists of about 1900
lines of code.
The \Abella proof script runs about 1600 lines. 
We note that \Abella has virtually no automation and the
current absence of polymorphism leads to some redundancy in the
proof.
We also note that, in contrast to Chlipala's work, our development
treats a version of the STLC that includes recursion. 
This necessitates the use of a step-indexed logical relation which
makes the overall proof more complex. 

Other frameworks have been proposed in the literature that facilitate
the use of \HOAS in implementing and verifying  compiler
transformations. 
Hickey and Nogin describe a framework for effecting compiler
transformations via rewrite rules that operate on a higher-order
representation of programs~\cite{hickey06hosc}. 
However, their framework is embedded within a
functional language. As a result, they are not able to support an
analysis of binding structure, an ability that brings
considerable benefit as we have highlighted in this paper.
Moreover, this framework offers no capabilities for verification. 
Hannan and Pfenning have discussed using a system called Twelf that is
based on LF in specifying and verifying compilers; see, for
example,~\cite{hannan92lics} and~\cite{murphy08modal} for some
applications of this framework. 
The way in which logical properties can be expressed in Twelf is
restricted; in particular, it is not easy to encode a logical 
relation-style definition within it.
The Beluga system~\cite{pientka10ijcar}, which implements a functional
programming language  based on contextual modal type
theory~\cite{nanevski08tocl}, overcomes some of the shortcomings of
Twelf.  
Rich properties of programs can be embedded in types in Beluga, and
Belanger \etal\ show how this feature can be exploited to ensure
type preservation for closure conversion~\cite{belanger13cpp}. 
Properties based on logical relations can also be described in
Beluga~\cite{cave15lfmtp}. 
It remains to be seen if semantics preservation proofs of the kind
discussed in this paper can be carried out in the Beluga system. 

While the framework comprising \LProlog and \Abella has significant
benefits in the verified implementation of compiler transformations
for functional languages, its current realization has some practical
limitations that lead to a larger proof development effort than seems
necessary.  
One such limitation is the absence of polymorphism in the Abella
implementation. 
A consequence of this is that the same proofs have sometimes to be
repeated at different types. 
This situation appears to be one that can be alleviated by allowing
the user to parameterize proofs by types and we are currently
investigating this matter.
A second limitation arises from the emphasis on explicit proofs
in the theorem-proving setup. 
The effect of this requirement is especially felt with respect to
lemmas about contexts that arise routinely in the $\lambda$-tree
syntax approach: such lemmas have fairly obvious proofs but,
currently, the user must provide them to complete the overall
verification task.
In the Twelf and Beluga systems, such lemmas are obviated by
absorbing them into the meta-theoretic framework.
There are reasons related to the validation of verification that lead
us to prefer explicit proofs. 
However, as shown in \cite{belanger14lfmtp}, it is often possible to
generate these proofs automatically, thereby allowing the user to
focus on the less obvious aspects.
In ongoing work, we are exploring the impact of using such ideas 
on reducing the overall proof effort. 

\subsubsection{Acknowledgements.}
We are grateful to David Baelde for his help in phrasing the
definition of the logical relation in Section~\ref{ssec:vericc}.  The
paper has benefited from many suggestions from its reviewers. This
work has been supported by the National Science Foundation grant
CCF-0917140 and by the University of Minnesota through a Doctoral
Dissertation Fellowship and a Grant-in-Aid of Research. Opinions,
findings and conclusions or recommendations that are manifest in this
material are those of the participants and do not necessarily reflect
the views of the NSF.

\newpage

\bibliographystyle{splncs03}
\bibliography{thisref}

\end{document}